\documentclass{article}
\usepackage{amsmath}
\usepackage{amsfonts}
\usepackage{amssymb}
\def\p{\partial}
\newtheorem{prop}{Proposition}

\newcommand{\dbar}{\bar{\partial}}
\newcommand{\wt}{\widetilde}
\newcommand{\be}{\begin{equation}}
\newcommand{\ee}{\end{equation}}
\newcommand{\bea}{\begin{eqnarray}}
\newcommand{\eea}{\end{eqnarray}}
\newcommand{\beaa}{\begin{eqnarray*}}
\newcommand{\eeaa}{\end{eqnarray*}}

\newcommand{\nn}{\nonumber}

\begin{document}
\title
{Doubrov-Ferapontov general heavenly equation and the hyper-K\"ahler hierarchy}
\author{
L.V. Bogdanov\thanks
{L.D. Landau ITP RAS,
Moscow, Russia}}
\date{}
\maketitle
\begin{abstract} We give a description of recently introduced Doubrov-Ferapontov
general heavenly equation in terms of closed differential Pl\"ucker two-form,
rationally depending on the spectral parameter. We demonstrate that general
heavenly equation is an important generating equation in the context 
of Takasaki hyper-K\"ahler hierarchy, and it is also  
directly connected to hyper-K\"ahler geometry
through the Gindikin construction. We develop a $\dbar$-dressing
scheme and introduce a formula for the potential satisfying
the general heavenly equation. Multidimensional generalization
is also outlined.
\end{abstract}
\section{Introduction}
General heavenly equation was introduced as a result of 
classification of integrable symplectic Monge-Amp\'ere equations in four dimensions \cite{DF}. It is one in the list of six equations,
and it is remarkably simple and symmetric, having the form
\bea
\alpha u_{12}u_{34} + \beta
u_{13}u_{24}+\gamma u_{14}u_{23} = 0,
\label{DF}
\eea
where $\alpha+\beta+\gamma$=0, subscripts denote partial derivatives.
The 
Lax pair was also presented in \cite{DF} in terms of 
vector fields $X_1$, $X_2$ in involution,
\bea
X_1=u_{34}\partial_1-u_{13}\partial_4+\gamma \lambda
(u_{34}\partial_1-u_{14}\partial_3), 
\nn\\
X_2=u_{23}\partial_4-u_{34}\partial_2+\beta \lambda
(u_{34}\partial_2-u_{24}\partial_3).
\label{DFpair}
\eea

In the present  work
we will give a description of Doubrov-Ferapontov
general heavenly equation (\ref{DF}), using the construction
developed in our works \cite{BK2013}, \cite{BK2014}, where we gave a formulation
of multidimensional dispersionless integrable hierarchy
in terms of differential $n$-form $\Omega$ in the space of
$N$ variables ($N\leqslant\infty$) $(x_0,\dots x_N)$,
possessing the following properties
\begin{enumerate}
\item The form $\Omega$ is decomposable, i.e.
\beaa
\Omega=\omega_1\wedge\dots\wedge\omega_n.
\eeaa
In algebraic terms that means that coefficients of the form
satisfy Pl\"uker relations, so we sometimes call it a Pl\"ucker form.
\item The form $\Omega$ is projectively closed, i.e., there exists
a function (gauge) $f(x_0,\dots x_N)$ such that
\beaa
d(f\Omega)=0.
\eeaa
\item The form $\Omega$ is projectively holomorphic with
respect to $x_0$, i.e., there exists a gauge $g(x_0,\dots x_N)$
such that coefficients of the form $g\Omega$ are holomorphic 
in some region of the complex plane of the variable
$\lambda=x_0$.
\end{enumerate}
The forms with these properties define multidimensional dispersionless
integrable hierarchy  in terms of integrable distribution of
holomorphic vector fields representing Lax operators
of the hierarchy. Two forms differing only by a gauge are equivalent
and define the same object. The case when it is possible to 
introduce the form $\Omega$ simultaneously holomorphic and closed in the standard sense
($f=g$) 
corresponds to important
reduction (preservation of volume), for which the basis of holomorphic
vector fields can be chosen divergence-free. Another important reduction
is HCR reduction, corresponding to heavenly equations and hyper-K\"ahler
hierarchies, for which the form $\Omega$ doesn't contain $d\lambda$
and $\lambda$ enters only parametrically.

To describe Doubrov-Ferapontov
general heavenly equation (\ref{DF}) in terms of this construction,
we will not need the most general version of the technique developed
in  \cite{BK2013}, \cite{BK2014}, because this equation belongs to HCR class,
and also it corresponds to the form $\Omega$ simultaneously holomorphic and closed in the standard sense 
(preservation of volume reduction).
We will demonstrate that general
heavenly equation (\ref{DF})
is an important generating equation in the context 
of hyper-K\"ahler hierarchy
\cite{Takasaki89}, \cite{Takasaki89a}.
We will also show  that it is 
directly connected to hyper-K\"ahler geometry and gives a solution
to complex
self-dual Einstein equation
through the Gindikin construction \cite{Gindikin85}, \cite{Gindikin86}.
\section{General heavenly equation through the differential two-form}
Let us consider 2-form depending on the spectral parameter
\bea
\Omega=\sum_{i,j} \omega_{ij}dx_i\wedge dx_j,
\label{2form}
\eea
where $1\leqslant i,j \leqslant 4$,
\bea 
\omega_{ij}(\lambda, \mathbf{x})=
\left(\frac{1}{\lambda-\lambda_i} - \frac{1}{\lambda-\lambda_j}\right)w_{ij}(\mathbf{x}),
\eea
$w_{ij}(\mathbf{x})$ is symmetric (here subscripts do not suggest differentiation).

Let $\Omega$ be a Pl\"ucker form. Pl\"ucker conditions 
for 2-forms are equivalent to the relation
\bea
\Omega\wedge\Omega=0,
\label{Plucker0}
\eea
which in our case gives one equation
\bea
\omega_{23}\omega_{14}-\omega_{13}\omega_{24}+
\omega_{12}\omega_{34}=0.
\eea
For $w_{ij}(\mathbf{x})$ we have
\bea
(\lambda_3-\lambda_2)(\lambda_4-\lambda_1)w_{23}w_{14}
-(\lambda_3-\lambda_1)(\lambda_4-\lambda_2)w_{13}w_{24}
\qquad &&\nn\\
+
(\lambda_2-\lambda_1)(\lambda_4-\lambda_3)w_{12}w_{34}=0.
&&
\label{Pluk}
\eea
Let us also suggest that $\Omega$ is closed,
\bea
\omega_{[ij,k]}=0.
\eea
For $w_{ij}(\mathbf{x})$ we have
\beaa
\frac{1}{\lambda-\lambda_2} (\p_1 w_{23}-\p_3 w_{12})+
\frac{1}{\lambda-\lambda_3} (\p_2 w_{13}-\p_1 w_{23})+
\frac{1}{\lambda-\lambda_1} (\p_3 w_{12}-\p_2 w_{13})=0,
\\
\frac{1}{\lambda-\lambda_2} (\p_1 w_{24}-\p_3 w_{12})+
\frac{1}{\lambda-\lambda_4} (\p_2 w_{14}-\p_1 w_{24})+
\frac{1}{\lambda-\lambda_1} (\p_3 w_{12}-\p_2 w_{14})=0,
\\
\frac{1}{\lambda-\lambda_4} (\p_1 w_{34}-\p_3 w_{14})+
\frac{1}{\lambda-\lambda_3} (\p_2 w_{13}-\p_1 w_{34})+
\frac{1}{\lambda-\lambda_1} (\p_3 w_{14}-\p_2 w_{13})=0,
\\
\frac{1}{\lambda-\lambda_2} (\p_1 w_{23}-\p_3 w_{24})+
\frac{1}{\lambda-\lambda_3} (\p_2 w_{34}-\p_1 w_{23})+
\frac{1}{\lambda-\lambda_4} (\p_3 w_{24}-\p_2 w_{34})=0.
\eeaa
These equations imply the existence of the potential 
$$
\Theta :
w_{ij}=\Theta_{,ij},
$$ 
and for arbitrary potential
$w_{ij}=\Theta_{,ij}$ satisfy the closedness equations.

Then Pl\"ucker relation (\ref{Pluk}) implies general heavenly
equation (\ref{DF}) for the potential,
\bea
(\lambda_3-\lambda_2)(\lambda_4-\lambda_1)\Theta_{,23}\Theta_{,14}
-(\lambda_3-\lambda_1)(\lambda_4-\lambda_2)\Theta_{,13}\Theta_{,24}
\qquad &&\nn\\
+
(\lambda_2-\lambda_1)(\lambda_4-\lambda_3)\Theta_{,12}\Theta_{,34}=0.
&&
\label{GHE}
\eea
\begin{prop}
\label{Prop}
Let us consider 2-form depending on the spectral parameter
\beaa
\Omega=\sum_{i,j}\left(\frac{1}{\lambda-\lambda_i} - \frac{1}{\lambda-\lambda_j}\right)
w_{ij}(\mathbf{x}) dx_i\wedge dx_j,
\eeaa
where $1\leqslant i,j \leqslant 4$,
$w_{ij}(\mathbf{x})$ is symmetric. The conditions
\beaa
\Omega\wedge\Omega&=&0,
\\
d\Omega&=&0
\eeaa
are equivalent to the existence of potential $\Theta$, $w_{ij}=\Theta_{,ij}$, satisfying the general heavenly 
equation
\beaa
(\lambda_3-\lambda_2)(\lambda_4-\lambda_1)\Theta_{,23}\Theta_{,14}
-(\lambda_3-\lambda_1)(\lambda_4-\lambda_2)\Theta_{,13}\Theta_{,24}
\qquad &&\nn\\
+
(\lambda_2-\lambda_1)(\lambda_4-\lambda_3)\Theta_{,12}\Theta_{,34}=0.
&&
\eeaa
\end{prop}
\section{Commuting flows and the `horizontal'
hierarchy}
It is possible not to restrict ourselves to the case
of four variables and consider the two-form (\ref{2form})
for arbitrary number of variables,
\bea
\Omega=\sum_{i,j=1}^N\left(\frac{1}{\lambda-\lambda_i} - \frac{1}{\lambda-\lambda_j}\right)
w_{ij}(\mathbf{x}) dx_i\wedge dx_j.
\label{multiform}
\eea
The closedness conditions for this two-form, in complete analogy with the
case of four variables,  imply the existence of the
potential $\Theta :
w_{ij}=\Theta_{,ij}$, and for every four distinct 
indices $1\leqslant i,j,k,l \leqslant N$ we have an equation
\bea
(\lambda_k-\lambda_j)(\lambda_l-\lambda_i)\Theta_{,jk}\Theta_{,il}
-(\lambda_k-\lambda_i)(\lambda_l-\lambda_j)\Theta_{,ik}\Theta_{,jl}
\qquad &&\nn\\
+
(\lambda_j-\lambda_i)(\lambda_l-\lambda_k)\Theta_{,ij}\Theta_{,kl}=0.
&&
\label{PlukGen}
\eea
Thus we have a kind of `horizontal' hierarchy of consistent
four-dimensional equations, where all the variables $x_i$
are on equal footing and correspond to simple poles. It
is possible to obtain general two-form meromorphic in
$\lambda$ by glueing simple poles of the form (\ref{multiform}). Moving this way, it is possible to
arrive to heavenly equation hierarchy \cite{Takasaki89}, where
the coefficients of the form are Laurent polynomials.

It is easy to include $\lambda_0=\infty$ into consideration
by the appropriate limit (we will denote the corresponding
variable $x_0$). The terms of two-form $\Omega$ containing
$d x_0$ read
\beaa
\Omega=\dots + 2\sum_{i=1}^N\frac{1}{\lambda-\lambda_i}
w_{i0}(\mathbf{x}) dx_i\wedge dx_0.
\eeaa
Equations (\ref{PlukGen}) containing partial derivative
over $x_0$ look like
\bea
(\lambda_k-\lambda_j)\Theta_{,jk}\Theta_{,i0}
-(\lambda_k-\lambda_i)\Theta_{,ik}\Theta_{,j0}
+
(\lambda_j-\lambda_i)\Theta_{,ij}\Theta_{,k0}=0.
&&
\label{PlukGen0}
\eea
\subsection{`Vacuum' two-form $\Omega$ and potential $\Theta$}
Let us start from a simple case of constant two-form $\Omega$,
\beaa
\Omega_0=\sum\left(\frac{1}{\lambda-\lambda_i} - \frac{1}{\lambda-\lambda_j}\right)
c_{ij} dx_i\wedge dx_j,
\eeaa
where $c_{ij}$ is constant and symmetric. The closedness 
condition is satisfied identically, and Pl\"ucker
condition (\ref{Pluk}) implies that the form $\Omega_0$
is decomposable,
\beaa
\Omega_0= 2 dS_0^1\wedge dS_0^2,
\eeaa
where
\bea
S_0^1=\sum_i\frac{a_i x_i}{\lambda-\lambda_i},\quad
S_0^2=\sum_i\frac{b_i x_i}{\lambda-\lambda_i},
\label{Svacuum}
\eea
and the constants $a_i$, $b_i$ satisfy the relations
\beaa
c_{ij}=\frac{a_ib_j-a_jb_i}{\lambda_i-\lambda_j}.
\eeaa
The `vacuum' potential $\Theta$ is quadratic in $x_i$,
\bea
\Theta_0=
\frac{1}{2}\sum_{i\neq j} 
\frac{a_ib_j-a_jb_i}{\lambda_i-\lambda_j}x_ix_j.
\label{Tvacuum}
\eea
For the general potential of the form 
\bea
\Theta=\Theta_0 + \tilde\Theta
\label{thetavac}
\eea
the terms entering the `vacuum' potential may be important
in the limit when we glue some of the points $\lambda_i$.

If we include $x_0$ corresponding to $\lambda_0=\infty$ into
consideration, for $S^1$, $S^2$, $\Theta_0$ we will have
additional terms
\bea
S^1_0=\dots + a_0 x_0, \quad S^2_0=\dots + b_0 x_0,
\label{Svacuum1}
\\
\Theta_0=\dots + \sum_{i} (a_ib_0-a_0 b_i)x_ix_0.
\nn
\eea
\section{From the horizontal hierarchy to the standard hyper-K\"ahler hierarchy}
General heavenly equation and the `horizontal hierarchy' connected with it
play the role of generating objects
for the heavenly equation hierarchy, or hyper-K\"ahler hierarchy 
\cite{Takasaki89,Takasaki89a},
which contains illustrious Pleba\'nski first and second heavenly equations and
higher equations. First, glueing simple poles of the two-form (\ref{multiform}),
it is possible to arrive to the general two-form with Laurent polynomial
coefficients, which corresponds to the heavenly equation hierarchy.
General heavenly equation (\ref{GHE}) plays a role of `dispersionless addition formula'
in this context. Substituting to it dispersionless vertex operators instead
of partial derivatives and taking into account vacuum terms of potential
$\Theta$, we get different generating equations
of the hierarchy. In this way it is possible, for example, to obtain
generating equations for the second heavenly equation hierarchy
introduced in \cite{BK2005,BK2006}.
\subsection{First heavenly equation from the general heavenly
equation}
First we discuss a simple example and demonstrate
how to obtain the first heavenly equation from the general heavenly
equation.
Let us consider a limit
\beaa
\lambda_1, \lambda_2\rightarrow \mu_1,
\quad
\lambda_3, \lambda_4\rightarrow \mu_2
\eeaa
for the potential $\Theta$ with some vacuum background
(\ref{thetavac}). First we pick out vacuum terms that
are singular in this limit,
$$
\Theta=\frac{a_1b_2-a_2b_1}{\lambda_1-\lambda_2}x_1x_2+
\frac{a_3b_4-a_4b_3}{\lambda_3-\lambda_4}x_3x_4 + \Theta'.
$$
Then, taking the limit, from equation (\ref{GHE}) 
for the function $\Theta'$ we get
\beaa
\Theta'_{,13}\Theta'_{,24}-
\Theta'_{,23}\Theta'_{,14}=
\frac{(a_1b_2-a_2b_1)(a_3b_4-a_4b_3)}
{(\mu_1-\mu_2)^2},
\eeaa
which is (up to a scaling) the  illustrious
Plebanski first heavenly equation.
Corresponding 2-form $\Omega$ is also obtained in this limit,
\bea
&&
\frac{1}{2}\Omega=
\frac{a_1b_2-a_2b_1}{(\lambda-\mu_1)^2}
dx_1\wedge d x_2+
\frac{a_3b_4-a_4b_3}{(\lambda-\mu_2)^2}
dx_3\wedge d x_4
\label{omegaH1}
\\
&&
\quad
+\frac{(\mu_1-\mu_2)\left(\Theta'_{,13}dx_1\wedge dx_3+
\Theta'_{,14}dx_1\wedge dx_4 +
\Theta'_{,23}dx_2\wedge dx_3 +
\Theta'_{,24}dx_2\wedge dx_4
\right)}{({\lambda-\mu_1})({\lambda-\mu_2})}.
\nn
\eea
Performing a M\"obius transformation of the spectral variable
$
\eta=\frac{\lambda-\mu_2}{\lambda-\mu_1},
$
for $\Omega$ (up to a factor) we get
\beaa
&&
\Omega\sim
\eta({a_1b_2-a_2b_1})
dx_1\wedge d x_2+
\frac{a_3b_4-a_4b_3}{\eta}
dx_3\wedge d x_4
\\
&&
\quad
+(\mu_1-\mu_2)\left(\Theta'_{,13}dx_1\wedge dx_3+
\Theta'_{,14}dx_1\wedge dx_4 +
\Theta'_{,23}dx_2\wedge dx_3 +
\Theta'_{,24}dx_2\wedge dx_4
\right).
\eeaa

Taking in (\ref{omegaH1}) $\mu_1=0$, $\mu_2=\infty$, $a_1=a_3=b_2=b_4=1$,
$a_2=a_4=b_1=b_3=0$, we get the two-form
\beaa
&&
\frac{1}{2}\Omega=
\frac{1}{\lambda^2}
dx_1\wedge d x_2+
dx_3\wedge d x_4
\\
&&
\qquad
+\frac{1}{\lambda}\left(\Theta'_{,13}dx_1\wedge dx_3+
\Theta'_{,14}dx_1\wedge dx_4 +
\Theta'_{,23}dx_2\wedge dx_3 +
\Theta'_{,24}dx_2\wedge dx_4
\right),
\eeaa
which corresponds to the standard setting for the first heavenly equation. Also for this case
\beaa
S^1_0=\frac{1}{\lambda}x_1+x_3,
\quad S^2_0=\frac{1}{\lambda}x_2+x_4.
\eeaa
\subsection{General heavenly equation and generating relations
for the second heavenly equation hierarchy}
Now we will demonstrate how to to obtain
generating equations for the second heavenly equation hierarchy
introduced in \cite{BK2005,BK2006} starting from the general
heavenly equation. Together with the times of `horisontal' hierarchy
(vertex times) we will consider standard infinite sets of times of
the second heavenly equation hierarchy,
\bea
S_0^1=\sum_i\frac{a_i x_i}{\lambda-\lambda_i} +
\sum_{n=0}^\infty t^1_n \lambda^n,
\quad
S_0^2=\sum_i\frac{b_i x_i}{\lambda-\lambda_i}+
\sum_{n=0}^\infty t^2_n \lambda^n.
\label{SvacuumH1}
\eea
Introducing vertex operators
\beaa
D^1(\mu)=-\sum_{n=0} \mu^{-(n+1)}\partial^1_n,
\quad 
D^2 (\mu)=-\sum_{n=0} \mu^{-(n+1)}\partial^2_n,
\eeaa 
where it is suggested that $|\mu|>1$,
we express derivatives over horizontal times through derivatives
over times $t^1_n$, $t^2_n$,
\bea 
\frac{\partial}{\partial x_i}=a_i D^1(\lambda_i)
+ b_i D^2(\lambda_i).
\label{vertex}
\eea 
Rewriting the general heavenly
equation (\ref{GHE}) for the function $\tilde\Theta$ (\ref{thetavac})
(which corresponds to Takasaki second key function \cite{Takasaki89}
and the `$\tau$-function' $\Theta$
for the second heavenly equation hierarchy of the work \cite{BK2005})
\beaa 
\Theta=\Theta_0 + \tilde\Theta,
\qquad
\Theta_0=
\frac{1}{2}\sum_{i\neq j} 
\frac{a_ib_j-a_jb_i}{\lambda_i-\lambda_j}x_ix_j,
\eeaa 
and substituting vertex expressions for derivatives (\ref{vertex}),
we get a generic generating relation for the second heavenly equation
hierarchy depending on four points $\lambda_1,\dots,\lambda_4$
and parameters $a_i$, $b_i$. Generating relations introduced
in \cite{BK2005,BK2006} contain three points and can be obtained by
glueing a pair of points. For example, let us consider a choice
\beaa 
S^1_0=\frac{x_1}{\lambda-\lambda_1}+
\frac{x_2}{\lambda-\lambda_2} +\dots,
\quad 
S^2_0=\frac{x_3}{\lambda-\lambda_3}+\frac{x_4}{\lambda-\lambda_4} +\dots.
\eeaa 
In this case
\beaa 
\Theta_0=\frac{x_1x_3}{\lambda_1-\lambda_3}+
\frac{x_1x_4}{\lambda_1-\lambda_4}+ 
\frac{x_2x_3}{\lambda_2-\lambda_3}+
\frac{x_2x_4}{\lambda_2-\lambda_4},
\eeaa 
and from (\ref{GHE}) for $\tilde\Theta$ we get
\bea 
(\lambda_2-\lambda_4)\tilde\Theta_{,24}
+(\lambda_3-\lambda_2)\tilde\Theta_{,23}
+(\lambda_4-\lambda_1)\tilde\Theta_{,14}
+(\lambda_1-\lambda_3)\tilde\Theta_{,13}
\qquad
\nn\\
=(\lambda_3-\lambda_2)(\lambda_4-\lambda_1)
\tilde\Theta_{,23}\tilde\Theta_{,14}
-(\lambda_3-\lambda_1)(\lambda_4-\lambda_2)
\tilde\Theta_{,13}\tilde\Theta_{,24}
\nn\\
+
(\lambda_2-\lambda_1)(\lambda_4-\lambda_3)
\tilde\Theta_{,12}\tilde\Theta_{,34}.
\label{gengen}
\eea 
Taking into account that expressions for derivatives through
vertex operators in this case are
\beaa 
\frac{\partial}{\partial x_1}=D^1(\lambda_1),\quad
\frac{\partial}{\partial x_2}=D^1(\lambda_2),\quad
\frac{\partial}{\partial x_3}=D^2(\lambda_3),\quad
\frac{\partial}{\partial x_4}=D^2(\lambda_4),
\eeaa
from (\ref{gengen}) we obtain a symmetric four-point 
generating relation for the second heavenly equation hierarchy.
Then, glueing $\lambda_1$ and $\lambda_4$, we get
a three-point generating relation
\beaa 
\frac{1}{\lambda_1-\lambda_3}
{D^1(\lambda_2)(D^2(\lambda_1)-D^2(\lambda_3))\tilde\Theta}
 - 
\frac{1}{\lambda_1-\lambda_2}
{D^2(\lambda_3)(D^1(\lambda_1)-D^1(\lambda_2))\tilde\Theta}
\\
=
D^1(\lambda_1)D^1(\lambda_2)\tilde\Theta
\cdot
D^2(\lambda_3)D^2(\lambda_1)\tilde\Theta
-
D^1(\lambda_1)D^2(\lambda_3)\tilde\Theta\cdot
D^1(\lambda_2)D^2(\lambda_1)\tilde\Theta,
\eeaa 
which is exactly one of the set of generating relations introduced
in \cite{BK2005,BK2006}; other generating relations can be obtained
in a similar way.
\section{Lax pair: vector fields in involution}
Here we use the technical setting described in 
\cite{BK2013}, \cite{BK2014}.  The two-form form $\Omega$ defines an associated 
subspace ${A}$  in the 
space of vector fields (distribution) defined by the
condition that interior product of vector field 
with the form is equal to zero,
$$
i_V
\Omega=0.
$$
The Pl\"ucker property (\ref{Plucker0}) 
(or decomposability of the form) guarantees that the 
dimension of this distribution
is exactly $(N-2)$, where $N$ is the number of variables.
The closedness of the Pl\"ucker form leads to involutivity of 
this distribution and the fact that basic vector fields 
can be chosen divergence-free.

Following \cite{BK2013}, it is easy to write down 
vector fields belonging to the
distribution $A$
associated with two-form $\Omega$ (\ref{multiform}) explicitly,
\beaa
U_{ijk}=\left( \frac{1}{\lambda-\lambda_i}-
\frac{1}{\lambda-\lambda_j}\right)w_{ij}\p_k
+\left( \frac{1}{\lambda-\lambda_j}-
\frac{1}{\lambda-\lambda_k}\right)w_{jk}\p_i
\\
+\left( \frac{1}{\lambda-\lambda_k}-
\frac{1}{\lambda-\lambda_i}\right)w_{ki}\p_j
\eeaa
Linear span of these vector fields  is 
$(N-2)$-dimensional in the tangent space (due to Pl\"ucker relations).
For the (projectively)
closed two-form $\Omega$ these vector fields are in involution. 
Divergence-free condition, implied by the standard closedness of the form $\Omega$,
is equivalent to 
the existence of potential $\Theta:
w_{ij}=\Theta_{,ij}$.

To find the Lax pair (a pair of vector fields in involution)
corresponding to Dubrov-Ferapontov general heavenly equation,
we consider a set of four indices, e.g., 1,2,3,4. Any pair of vector fields
$U_{ijk}$ with distinct $i,j,k$ belonging to our set is in 
involution and constitutes a Lax pair.

In equivalent form, taking $U_{ijk}\rightarrow (\lambda-\lambda_i)(\lambda-\lambda_j)(\lambda-\lambda_k)U_{ijk}$,
we get polynomial fields of the first order in spectral parameter:
\beaa
U_{ijk}=(\lambda_i-\lambda_j)\Theta_{,ij}(\lambda-\lambda_k)\p_k+
(\lambda_j-\lambda_k)\Theta_{,jk}(\lambda-\lambda_i)\p_i+
\\
(\lambda_k-\lambda_i)\Theta_{,ki}(\lambda-\lambda_k)\p_j.
\eeaa
After M\"obius transformation 
of the spectral variable, it is possible to get the
Lax pair exactly in Doubrov-Ferapontov form (\ref{DFpair}).
\subsection{Associated system of one-forms}
Associated system of one-forms for $\Omega$ is a 
linear subspace $A^*$ in the space of one-forms
dual to distribution $A$,
$$
\langle A^*, A\rangle=0.
$$
Due to Pl\"ucker relations, this subspace is 2-dimensional
(locally in cotangent space),
$\Omega$ is decomposable and can be represented as
\beaa
\Omega=\psi\wedge\phi,
\eeaa
where $\psi,\phi\in A^*$. For arbitrary  vector field
$V$ 
\beaa
i_V \Omega\in A^*.
\eeaa 
Taking $V_p=\partial_p$, we get
the following one-forms belonging to $A^*$,
\beaa
\phi_p:=i_{V_p}\Omega=\sum_{i:\; i\neq p}^N
\left(\frac{1}{\lambda-\lambda_i}-
\frac{1}{\lambda-\lambda_p}\right) w_{pi}d x_i.
\eeaa
For the four-dimensional case ($N$=4),
it is possible to construct 
the basis of polynomial forms of the first order
in $\lambda$, e.g.
\bea
\phi^1_{34}= (\lambda-\lambda_2)(\lambda-\lambda_3)(\lambda-\lambda_4)
(w_{14}\phi_3-w_{13}\phi_4),
\nn\\
\phi^2_{34}= (\lambda-\lambda_1)(\lambda-\lambda_3)(\lambda-\lambda_4)
(w_{24}\phi_3-w_{23}\phi_4).
\label{1formsbasis}
\eea
This basis is important to establish a correspondence of 
Doubrov-Ferapontov general heavenly equation with Gindikin construction.
\section{Gindikin construction}
The original statement from Gindikin work \cite{Gindikin85} reads (citation, 
translated from the Russian text):

\textit{Construction of complex solutions of self-dual
Einstein equation is equivalent to
construction of quadratic (in $t$) bundles of holomorphic two-forms
$F(t)=t^2F_2+tF_1+F_0$, $t\in\mathbb{C}$, on the four-dimensional complex
manifold $M$, satisfying the conditions
\bea
(i)~~~\qquad F(t)\wedge F(t)&=&0\quad \text{for all $t$};
\nn\\
(ii)~~~~~~~~~\qquad d F(t)&=&0;
\\
(iii)~\qquad F(t)\wedge F(s)&\neq& 0\quad \text{for $t\neq s$.}
\nn 
\eea
Condition (iii) means nondegeneracy and it is often convenient
to ignore it in the process of calculations.
Due to condition (i) $F(t)$ can be represented as
\bea
F(t)=(\phi_0+t\phi_1)\wedge (\psi_0+t\psi_1),
\eea
where $\phi_i$, $\psi_i$ are 1-forms. Then condition (iii) guarantees
non-degeneracy of the metric
\bea
g=\phi_0\psi_1-\phi_1\psi_0,
\label{Gmetric}
\eea
and condition (ii) implies that it is right-flat  (satisfies self-dual Einstein equation).}

Let us consider the form
\bea
F(\lambda)=(\lambda-\lambda_1)(\lambda-\lambda_2)
(\lambda-\lambda_3)(\lambda-\lambda_4)\Omega,
\label{formF}
\eea
where
\beaa
\Omega=\sum_{1\leqslant i,j \leqslant 4}
\left(\frac{1}{\lambda-\lambda_i} - \frac{1}{\lambda-\lambda_j}\right)
\Theta_{,ij}(\mathbf{x}) dx_i\wedge dx_j,
\eeaa
and $\Theta$ satisfies the general heavenly equation
(\ref{GHE}). 
The form $F(\lambda)$ is quadratic in $\lambda$ and,
due to Prop. \ref{Prop}, satisfies the conditions required by Gindikin's statement.

The metric can be constructed explicitly, using the basis
(\ref{1formsbasis}), which reads
\beaa
\phi_{34}^1&=&
(\lambda-\lambda_2)(\lambda_3-\lambda_4)\Theta_{,14}\Theta_{,13}dx_1
\\&&+
((\lambda-\lambda_4)(\lambda_3-\lambda_2)\Theta_{,23}\Theta_{,14}
-(\lambda-\lambda_3)(\lambda_4-\lambda_2)\Theta_{,24}\Theta_{,13})dx_2
\\&&+
(\lambda-\lambda_2)(\lambda_3-\lambda_4)\Theta_{,13}\Theta_{,34}dx_3
\\&&+
(\lambda-\lambda_2)(\lambda_3-\lambda_4)\Theta_{,14}\Theta_{,34}dx_4,
\eeaa
\beaa
\phi_{34}^2&=&
((\lambda-\lambda_4)(\lambda_3-\lambda_1)\Theta_{,13}\Theta_{,24}
-(\lambda-\lambda_3)(\lambda_4-\lambda_1)\Theta_{,14}\Theta_{,23})dx_1
\\&&+
(\lambda-\lambda_1)(\lambda_3-\lambda_4)\Theta_{,24}\Theta_{,23}dx_2
\\&&+
(\lambda-\lambda_1)(\lambda_3-\lambda_4)\Theta_{,23}\Theta_{,34}dx_3
\\&&+
(\lambda-\lambda_1)(\lambda_3-\lambda_4)\Theta_{,24}\Theta_{,34}dx_4.
\eeaa
It is easy to check that the form $F$ is expressed
through this basis as
\beaa
F=\frac{2 \phi_{34}^1 \wedge \phi_{34}^2}
{(\lambda_3-\lambda_4)(\Theta_{,13} \Theta_{,24}  
-\Theta_{,14} \Theta_{,23})\Theta_{,34}  
}
\eeaa
The metric $g$ is then given by the formula
\beaa
g=
\frac{2(\phi_{34(0)}^1 \phi_{34(1)}^2-
\phi_{34(1)}^1 \phi_{34(0)}^2)}
{(\lambda_3-\lambda_4)(\Theta_{,13} \Theta_{,24}  
-\Theta_{,14} \Theta_{,23})\Theta_{,34}  
}
\eeaa
where by the subscripts (0), (1) we denote the terms
of the zero and first order with respect to $\lambda$.
Explicit expressions for the components of the metric are
\bea
&&
g_{ii}=2G \Theta_{,ij}\Theta_{,ik}\Theta_{,ip},
\nn\\&&
g_{kp}={G}\Theta_{,kp}(\Theta_{,ik}\Theta_{,jp}
+\Theta_{,ip}\Theta_{,jk}
),
\label{metric}
\eea
where $i,j,k,p$ are pairwise distict,
\beaa
G=\frac{(\lambda_i-\lambda_j)(\lambda_k-\lambda_p)}{  
\Theta_{,ik} \Theta_{,jp} -\Theta_{,ip} \Theta_{,jk} }.
\eeaa
It is easy to check that the expression for $G$ is 
invariant under arbitrary
permutation of indices due to the general heavenly equation.
Thus, starting from a solution of the general 
heavenly equation (\ref{GHE}),
via Gindikin method  we have constructed a (complex) metric 
(\ref{metric})
satisfying
self-dual Einstein equations. 

It is interesting to note that the work
\cite{FK} gives a direct recipe to calculate a 
self-dual conformal structure for equations
of the heavenly type through the symbol of linearization of equation,
which represents a symmetric bivector defining a conformal structure.
For the general heavenly equation (\ref{GHE}) this bivector reads
\beaa 
\gamma^{ij}=\epsilon^{ijkp}
(\lambda_i-\lambda_j)(\lambda_k-\lambda_p)\Theta_{,kp},
\eeaa 
and it is easy to check that inverse matrix to $\gamma^{ij}$
gives the metric $g_{ij}$ (\ref{metric}) (up to a factor),
thus the conformal structure is the same as in Gindikin
construction. This natural conjecture belongs to E.V. Ferapontov 
and it can be proved for the general case using the representation
of the symbol of linearization in terms of the basis of
vector fields of the first order in $\lambda$ \cite{FK}, which is
in some sense dual to representation (\ref{Gmetric}),
and results
of the works \cite{Dun99}, \cite{GS99}.
However, to get self-dual
metric satisfying the Einstein equation
(Ricci-flat), it is important to define the normalization,
because this property is not conformally-invariant,
and Gindikin construction provides a direct answer to this question
(\ref{metric}).

\section{$\dbar$-dressing scheme}
In this section we use the technique developed in \cite{BK2005}
in the context of Pleba\'nski second heavenly equation hierarchy.
Due to the fact that main results were formulated in variational form,
they are applicable to our present setting with minor modifications.

The two-form $\Omega$ can be represented as
\beaa
\Omega=dS^1\wedge dS^2.
\eeaa
Two properties of $\Omega$ are now identically satisfied 
(it is Pl\"ucker and closed), now the problem is to construct
$S^1$, $S^2$ to get $\Omega$ with the necessary analytic properties.

Let us consider
nonlinear vector $\dbar$ problem in some region $G$,
\bea
\dbar S^1&=&W_{,2}(\lambda,\bar \lambda;S^1,S^2), 
\quad W_{,2}:=\frac{\p W}{\p S^2}
\nn\\
\dbar S^2&=&-W_{,1}(\lambda,\bar \lambda;S^1,S^2), \quad W_{,1}:=\frac{\p W}{\p S^1}.
\label{dbar}
\eea
and $W(\lambda,\bar \lambda;S^1,S^2)$ is some function defined in $G$.
This problem provides analyticity of the form 
$\Omega=dS^1\wedge dS^2$ in $G$.

We search for solutions of the form
\beaa
S^1=S^1_0+ \tilde S^1,\quad S^2=S^2_0+ \tilde S^2
\eeaa
where $\tilde S^1$, $\tilde S^2$ are analytic outside $G$
and go to zero at infinity,
$S^1_0$, $S^2_0$ are analytic in $G$ (normalization or vacuum term, compare
(\ref{Svacuum}), (\ref{Svacuum1}))
\beaa
S^1_0=\sum_{i=1}^{N} \frac{a_i x_i}{\lambda-\lambda_i}, 
\quad S^2_0=\sum_{i=1}^{N} \frac{b_i x_i}{\lambda-\lambda_i}.
\eeaa
Then the form $\Omega$ has the required analytic structure.

The $\dbar$ problem can be obtained by variation of the action
\bea
f
=\frac{1}{2\pi\mathrm{i}}\iint_{G}
\left(\wt S^2 \dbar \wt S^1 -
W(\lambda,\bar \lambda,S^1,S^2)\right)d\lambda\wedge d\bar \lambda,
\label{action}
\eea
where one should consider independent variations of $\wt{\mathbf{S}}$,
possessing required analytic properties, 
keeping $\mathbf{S}_0$ fixed. Using the results of the work 
\cite{BK2005} in our setting, we come to the following statement:
\begin{prop} 
The function
\bea
\Theta(\mathbf{x})=\Theta_0+
\frac{1}{2\pi\mathrm{i}}\iint_{G}
\Bigl(\wt S^2(\mathbf{x}) \dbar \wt S^1(\mathbf{x})
-
W(\lambda,\bar \lambda,S^1(\mathbf{x}),S^2(\mathbf{x}))\Bigr) 
d\lambda\wedge d\bar \lambda,
\label{HEtau}
\eea
where $\Theta_0$ is a vacuum term defined by formula (\ref{Tvacuum}),
\bea
\Theta_0=
\frac{1}{2}\sum_{i\neq j} \frac{a_ib_j-a_jb_i}
{\lambda_i-\lambda_j}x_ix_j.
\label{Tvacuum1}
\eea
i.e., the action (\ref{action}) evaluated on the solution
of the $\dbar$ problem (\ref{dbar}) plus a 
term quadratic in $x_i$,
is a solution of the hierarchy of 
Doubrov-Ferapontov general heavenly equations
(\ref{PlukGen}).
\end{prop}

A class of solutions of the general heavenly equation hierarchy
(\ref{PlukGen})
in terms of implicit functions (similar to \cite{Gindikin85},
\cite{Takasaki89})
can be constructed using the choice
\beaa
W(\lambda,\bar \lambda,S^1,S^2)=
2\pi\mathrm{i}\left(\sum_{i=1}^{M}\delta(\lambda-\mu_i)F_i(S^1)
+
\sum_{i=1}^{M}\delta(\lambda-\nu_i)G_i(S^2)\right),
\eeaa
where $\delta(\lambda-\mu_i)$, $\delta(\lambda-\nu_i)$ are two-dimensional
delta functions in the complex plane, and $F_i$, $G_i$ are some functions of
one variable. The $\dbar$ problem (\ref{dbar}) in this case reads
\bea
\dbar \tilde S^1&=&2\pi\mathrm{i}\sum_{i=1}^{M}\delta(\lambda-\nu_i)G'_i(S^2)
\nn\\
\dbar \tilde S^2&=&-2\pi\mathrm{i}\sum_{i=1}^{M}\delta(\lambda-\mu_i)F'_i(S^1).
\label{dbardelta}
\eea
The solutions of the $\dbar$ problem are then of
the form
\beaa
\tilde S^1=\sum_{i=1}^{M} \frac{f_i}{\lambda-\nu_i}, \quad  
\tilde S^2=\sum_{i=1}^{M} \frac{g_i}{\lambda-\mu_i},
\eeaa
and from (\ref{dbar}) the functions $f_i$, $g_i$ are defined as implicit functions,
\bea
f_i(\mathbf{x})&=&G'_i\left(\sum_{j=1}^{N} \frac{b_j x_j}
{\nu_i-\lambda_j}+
\sum_{k=1}^{M} \frac{g_k(\mathbf{x})}{\nu_i-\mu_k}\right),
\nn\\
g_i(\mathbf{x})&=&-F'_i\left(\sum_{j=1}^{N} \frac{a_j x_j}
{\mu_i-\lambda_j}+
\sum _{k=1}^{M}\frac{f_k(\mathbf{x})}{\mu_i-\nu_k}\right).
\label{impl}
\eea 
The potential $\Theta$ solving the general heavenly equation hierarchy
is then given by the formula (\ref{HEtau}), it depends on the set
of arbitrary functions of one variable $F_i$, $G_i$,
\bea 
\Theta(\mathbf{x})=\Theta_0 + 
\sum_{i=1}^{M} F_i(S^1(\mu_i))
+
\sum_{i=1}^{M} G_i(S^2(\nu_i))+
\sum_{i=1}^{M}\sum_{j=1}^{M} \frac{f_i g_j}{\nu_i - \mu_j},
\label{Theta}
\eea 
where $\Theta_0$ is given by (\ref{Tvacuum1}),
\beaa 
S^1=S^1_0+\tilde S^1=\sum_{i=1}^{N}
\frac{a_i x_i}{\lambda-\lambda_i}+
\sum_{i=1}^{M} \frac{f_i}{\lambda-\nu_i},
\\
S^2=S^2_0+\tilde S^2=\sum_{i=1}^{N}
\frac{b_i x_i}{\lambda-\lambda_i}+
\sum_{i=1}^{M} \frac{g_i}{\lambda-\mu_i},
\eeaa 
and functions $f_i$, $g_i$ are defined as implicit functions 
by equations (\ref{impl}). Formula (\ref{Theta}) corresponds
to the special solution of hyper-K\"ahler hierarchy derived
in \cite{Takasaki89}.
\section{On the multidimesional hyper-K\"ahler case}
We will briefly outline the formulation of multidimensional case,
which is mostly similar to the four-dimensional case discussed
above.
Let us consider the two-form $\Omega$ of the same structure (\ref{2form}), 
but now satisfying the conditions
\bea
\Omega\wedge\dots\wedge\Omega&=&0\quad (N\text{~times}),
\label{N}\\
d\Omega&=&0,
\nn
\eea
that correspond to the setting for multidimensional
hyper-K\"ahler case considered in \cite{Gindikin86}, \cite{Takasaki89}.
In terms of construction of the works \cite{BK2013}, 
\cite{BK2014}, the basic
decomposable (Pl\"ucker) form is $\widetilde \Omega=
\Omega\wedge\dots\wedge\Omega$ ($N-1$ times), and 
the multidimesional hyper-K\"ahler case is a reduction of the general case.

Closedness of $\Omega$, as in the four-dimensional case, is  
equivalent to the existence of potential $\Theta$.
Then from relation (\ref{N}) for every set
of $2N$ distinct indices $i_1,\dots,i_{2N}$
we obtain
$2N$-dimensional homogeneous equation of degree $N$,
which may be considered `general hyper-K\"ahler equation'
\beaa
\sum\epsilon_{i_1\dots i_{2N}}(\lambda_{i_1}-\lambda_{i_2})
\times\dots\times
(\lambda_{i_{2N-1}}-\lambda_{i_{2N}})
\Theta_{,i_1 i_2}\times\dots\times\Theta_{,i_{2N-1} i_{2N}}=0,
\eeaa
where summation is over permutation of indices. This equation is
a generating equation for multidimensional
hyper-K\"ahler hierarchy \cite{Takasaki89}.

The form $\Omega$ can be represented as
\beaa
\Omega=S^1\wedge S^2+\dots+S^{2N-3}\wedge S^{2N-2}.
\eeaa 
Similar to the four-dimensional case, 
it is possible to formulate $\dbar$-dressing scheme and
find a
formula for $\Theta$ completely analogous to (\ref{HEtau}). 
\section*{Acknowledgements}
This research  
was partially supported  by RFBR grant 14-01-00389. The author
is grateful to E.V. Ferapontov for useful discussions.

\end{document}